\def\beg{\begin{equation}}
\def\eeq{\end{equation}}
\documentstyle[12pt]{article}
\begin{document}
\begin{center}
{\Large{\bf Comments on ``Evidence of Landau Levels and Interactions
in Low-Lying Excitations of Composite Fermions..." by Dujovne, Pinczuk, 
Kang, Dennis, Pfeiffer and West, cond-mat/0211022.}}
\vskip0.35cm
{\bf Keshav N. Shrivastava}
\vskip0.25cm
{\it School of Physics, University of Hyderabad,\\
Hyderabad  500046, India}
\end{center}

Dujane et al suggest that observed spectra are a result of 
spin-split Landau levels and spin-flip energies reveal composite fermion(CF)interactions. We find that the CF model is independent
of spin so that the interpretations of data by Dujovne et al in 
terms of CF model are incorrect. It may be pointed out that the experimental mass of the quasiparticles is several orders of 
magnitude smaller than the CF mass.
\vfill
Corresponding author: keshav@mailaps.org\\
Fax: +91-40-3010145.Phone: 3010811.
\newpage
\baselineskip22pt
\noindent {\bf 1.~ Introduction}

     Laughlin$^1$ has found a wave function which is antisymmetric 
for odd values of an integer, $m$. Afterwords, they have corrected
their paper to include even values of $m$ which give bosons$^2$.
Similarly, even numerators with odd denominators give bosons. For
odd $m$, the square of the wave function describes a system of charge density,
\beg
\sigma_m=1/[m(2\pi a_o^2)].
\eeq
Laughlin has argued that this means that the charge has become 
$e/m$. However, Laughlin has not considered the possibility of 
keeping $e$ unchanged and change $a_o^2$ to $ma_o^2$. Clearly, there 
are three possibilities, (1) change $e$ to $e/m$ or (2) keep $e$ unchanged and change $a_o^2$ to $ma_o^2$. (3) There is, of course, 
a third possibility in which neither $e$ nor $a_o^2$ are changed 
and the integer $m$ just remain as a multiplier. Laughlin has 
considered only the first possibility so that his results are not 
unique as pointed out by us$^3$. Here $2\pi a_o^2$=hc/eH. Therefore, when $a_o$ is changed to ${\sqrt m} a_o$, five possibilities occur, i.e., to change any one of (i)h, (ii)c, (iii)e, (iv)H or (v) none.
Again, it is clear that throwing the entire blame on $e$ alone is 
not justified. The antisymmetry of wave function can be obtained 
for $m=1$ and $m=3$ is unnecessary so that there is no particular 
reason to consider $m=3$ which gives $e/3$ as the charge.

     Another model called composite fermion (CF) model has been suggested by Jain$^{4,5}$ in 1989-1998. According to this model, 
even number of flux quanta are attached to one electron. The even number was chosen to give the experimentally found odd denominator. The ``even number" gives the fermions and odd number of flux quanta give the bosons. CF is content with ``even number" of flux quanta attachment to the electrons. Therefore, all quasiparticles in the CF model must be fermions to be consistent with ``even number" of flux quanta attachment. However, the 2/3 is a boson and it is found in the data. Therefore, the CF model is internanally inconsistent$^6$.

     The experimentalists have used the CF model very extensively because it has the correct denominators. In the present comment, it is pointed out that the experimental data has nothing to do with the CF model. It is all the more important to note that the CF model is internally inconsistent and hence should be discarded. What is meant is that the {\it``flux quanta" attached to the electron} is not correct, theoretically.

\noindent{\bf 2.~~Comments.}

\noindent(i) Dujovne et al (page 1, left column, bottom) state that 
``spin-reversed quasiparticle-quasihole pairs have spin flip energies that are strongly affected by residual CF interactions. States such as 2/5 ... CF Landau level".

We wish to comment that CF model is independent of spin and the 
fraction 2/5 is not a fermion and hence not a CF.

\noindent(ii) Page 1, column 2, bottom: `` At filling factors close 
to 2/5 we observe modes due to spin-flip (SF) transitions 1$\uparrow
 \to$ 0 $\downarrow$ in which there are simultaneous changes in spin and 
CF Landau level quantum numbers, as shown in the inset to Fig.2(b)"

In fact this is not a feature of the CF model. This kind of spin-flip occurs only in Shrivastava's paper which is dated three years before 
CF model. The sequence in which discoveries have been made and the assignment of credits by Dujovne et al is incorrect. This kind of 
spin flip found in the data does not belong to CF model.

\noindent(iii) Page 2, column 2: ``Fig.2(b) shows a calculated dispersion of SF excitations at 2/5".

The 2/5 is not a fermion and hence not a CF. The CF model does not 
have rotations. Therefore any calculation, if agrees with data, is 
at best fortituous. The Fig.2(b) shows even flux attachment which 
makes them fermions whereas 2/5 is not a fermion. Similarly, the interpretation given in Fig.3(a) is incorrect.

\noindent(iv)Page 4, left column: The mass of the CF is found to be 0.4m$_o$. Actually, due to flux quanta attachment, the mass of the 
CF will be several hundred times the mass of the electron. 
Therefore, the experimentally measured mass is not in accord with 
the CF model by a few orders of magnitude. It may be mentioned that 
CF are large objects so that they can not be accomodated within the given space with the same density as that of the electrons. 
Therefore CF is a hypothetical object not found in nature and 
certainly not in GaAs.

\noindent{\bf3. ~~ Prizes and awards}

     Laughlin has been awarded half of the 1998 Nobel prize for 
writing {\bf the gound state wave function, the excitations of 
which have charge 1/3}. However, the factor of 1/3 can be adjusted
elsewhere so that the charge is not uniquely determined. Similarly,
the antisymmetry of the wave function may be satisfied with integer,
m=1 and m=3 is not necessary. What determines the antisymmetry
of the wave function does not necessarily determine the charge 
because there are other candidates to absorb the integer, m.
Similarly, the 2002 Oliver Buckley Prize of the APS was awarded 
to Jain for composite fermions (CF) but these CF are unphysical 
objects not found in GaAs.

\noindent{\bf4.~~ Conclusions}.

   Dujovne et al's experimental data has nothing to do with composite fermion (CF) model. They have incorrectly assigned the data to the CF model. We have pointed out that CF model is internally inconsistent$^{7,8}$. Dyakonov$^9$ has shown that CF model has no theoretical basis. Farid$^{10}$ has shown that CF field attachment formula is incorrect.

Ref.3 shows that the fractional charge need not arise from 
Laughlin's wave functions. Ref.6 shows that the CF model is 
unphysical. Ref.7 shows that CF model violates the principles of classical electrodynamics. Ref.8 shows that CF model is internally inconsistent. Ref.9 shows that the CF does not have a theoretical 
basis. Ref.10 shows that the field formula of CF is not correct.
Ref.11 shows that there are alternatives to CF model which agree 
with the data. Ref.12 shows that CF features do not match with the 
data. Ref. 13 shows that experimental data has been incorrectly described.

     The correct theory of the quantum Hall effect is given in ref.14.
\newpage

\noindent{\bf6.~~References}
\begin{enumerate}
\item R. B. Laughlin, Phys. Rev. Lett. {\bf50}, 1395(1983).
\item V. Kalmeyer and R. B. Laughlin, Phys. Rev. Lett. {\bf59}, 2095(1987).
\item K. N. Shrivastava, cond-mat/0210238.
\item J. K. Jain, Phys. Rev. Lett. {\bf63}, 199(1989).
\item K. Park and J. K. Jain, Phys. Rev. Lett.{\bf80}, 4237(1998).
\item K. N. Shrivastava, cond-mat/0210320.
\item K. N. Shrivastava, cond-mat/0209666.
\item K. N. Shrivastava, cond-mat/0209057.
\item M. I. Dyakonov, cond-mat/0209206.
\item B. Farid, cond-mat/0003064.
\item K. N. Shrivastava, cond-mat/0207391.
\item K. N. Shrivastava, cond-mat/0204627.
\item K. N. Shrivastava, cond-mat/0202459.
\item K. N. Shrivastava, Introduction to quantum Hall effect,\\ 
      Nova Science Pub. Inc., N. Y. (2002).
\end{enumerate}
\vskip0.1cm
Note: Ref.14 is available from:\\
 Nova Science Publishers, Inc.,\\
400 Oser Avenue, Suite 1600,\\
 Hauppauge, N. Y.. 11788-3619,\\
Tel.(631)-231-7269, Fax: (631)-231-8175,\\
 ISBN 1-59033-419-1 US$\$69$.\\
E-mail: novascience@Earthlink.net

\vskip0.5cm

\end{document}